\documentclass[aps,prl,reprint,superscriptaddress]{revtex4-1}
\pdfoutput=1
\usepackage{easy-todo}
\usepackage{color}
\usepackage{graphicx}
\usepackage{amsmath}
\usepackage{amsfonts}
\usepackage{amssymb}
\usepackage{microtype}
\usepackage{verbatim}
\usepackage{bbold}
\usepackage{hyperref}
\usepackage{cleveref}
\allowdisplaybreaks
\usepackage{multirow}
\usepackage{bbm}
\usepackage{braket}
\hypersetup{
  colorlinks   = true, 
  urlcolor     = blue, 
  linkcolor    = blue, 
  citecolor   = blue 
}

\begin{document}

\author{Pieter W. Claeys}
\email{claeys@pks.mpg.de}
\affiliation{Max Planck Institute for the Physics of Complex Systems, 01187 Dresden, Germany}

\author{Giuseppe De Tomasi}
\email{detomasi@illinois.edu}
\affiliation{Max Planck Institute for the Physics of Complex Systems, 01187 Dresden, Germany}
\affiliation{Department of Physics and Institute for Condensed Matter Theory, University of Illinois at Urbana-Champaign, Urbana, Illinois 61801-3080, USA}

\title{Fock-space delocalization and the emergence of the Porter-Thomas distribution \\ from dual-unitary dynamics}

\begin{abstract}
The chaotic dynamics of quantum many-body systems are expected to quickly randomize any structured initial state, delocalizing it in Fock space. In this work, we study the spreading of an initial product state in Hilbert space under dual-unitary dynamics, captured by the inverse participation ratios and the distribution of overlaps (bit-string probabilities). We consider the self-dual kicked Ising model, a minimal model of many-body quantum chaos which can be seen as either a periodically driven Floquet model or a dual-unitary quantum circuit. Both analytically and numerically, we show that the inverse participation ratios rapidly approach their ergodic values, corresponding to those of Haar random states, and establish the emergence of the Porter-Thomas distribution for the overlap distribution. Importantly, this convergence happens exponentially fast in time, with a time scale that is independent of system size. We inspect the effect of local perturbations that break dual-unitarity and show a slowdown of the spreading in Fock space, indicating that dual-unitary circuits are maximally efficient at preparing random states. 
\end{abstract}

\maketitle

\emph{Introduction.---}
Understanding the emergence of statistical mechanics from the out-of-equilibrium dynamics of quantum many-body systems is an active research front with many potential applications in quantum technologies.  
Generic many-body quantum systems are believed to thermalize under their own unitary time evolution, where the long-time quantum expectation value of a local observable matches its thermodynamic value as specified by a few global conserved quantities. Any local structure of initial states that is not encoded in such conserved quantities is effectively erased under chaotic dynamics~\cite{Nandkishore_review_2015,dalessio_quantum_2016}. From a different perspective, thermalization is deeply connected to the notion of ergodicity in the Hilbert space, which implies an equipartition of the many-body wave function over the available many-body Fock states~\cite{Guhr_review_1998, Borgonovi_review_2016}. Initial wave functions are expected to spread and delocalize in Fock space, approaching a long-time state that is indistinguishable from a (Haar) random state.
This emergent randomness underlies fundamental quantum notions of thermalization, scrambling, and entanglement growth~\cite{Nandkishore_review_2015,dalessio_quantum_2016,Guhr_review_1998, Borgonovi_review_2016,Altshuler_1997,Gornyi_2005,BASKO_2006,nahum_operator_2018,von_keyserlingk_operator_2018,khemani_operator_2018,hunter-jones_unitary_2019,de_tomasi_multifractality_2020,GDT_2021,fisher2023random,pappalardi_general_2023,fava_designs_2023}, and the resulting Haar random states, in turn, present a valuable practical resource for quantum information, quantum tomography, and simulation~\cite{Nielsen_Chuang_2010,huang_predicting_2020,richter_simulating_2021,mcginley_shadow_2023,mark_benchmarking_2023}.

Indeed, Hilbert space delocalization has very recently gained intense interest in the context of quantum computing, being the backbone of quantum random sampling -- the leading test of quantum advantage~\cite{boixo_characterizing_2018,arute_quantum_2019,hangleiter_computational_2023}. Here the fundamental object of interest is the probability distribution of the Fock-space amplitudes, returning the celebrated Porter-Thomas distribution for Haar random states~\cite{Porter_1956,haake2001quantum,mullane_sampling_2020}. This distribution more generally serves as a diagnostic for maximum entropy principles~\cite{mark_maximum_2024}.

In the context of quantum dynamics, the understanding of this delocalization is largely built on numerical evidence, with rigorous analytical results being scarce (away from the random circuit case~\cite{fisher2023random,turkeshi_hilbert_2024,bertoni_shallow_2024,dalzell_random_2022,fefferman_anti-concentration_2024,christopoulos_universal_2024}). At late times the Porter-Thomas distribution was proven to emerge under mild assumptions~\cite{mark_benchmarking_2023,mark_maximum_2024}, where the necessary time scales and intermediate dynamics are however not known. Ref.~\cite{turkeshi_hilbert_2024} showed that under random unitary circuits the participation entropies of a many-body state approach late-time values consistent with Haar-random states in a time that scales logarithmically with system size. 

In this work, we present a full analytic characterization of the spreading in Fock space of an initially localized state under dual-unitary dynamics, i.e., dynamics that are unitary in both space and time. 
We specifically consider the self-dual kicked Ising model, a paradigmatic model of chaotic many-body dynamics. We inspect the finite-time dynamics and prove Fock-space ergodicity by showing that the overlap distribution exponentially approaches the Porter-Thomas distribution as time increases. Interestingly, we find that the time scale of this approach is independent of system size, to be contrasted with the random circuit result~\cite{turkeshi_hilbert_2024}. 
Moving away from the dual-unitary point, introducing a dual-unitarity-breaking perturbation is shown to slow down spreading.
These results provide evidence that dual-unitary circuits are maximally efficient at preparing random states.

\emph{Model $\&$ methods}.---
We consider the dynamics of a one-dimensional chain of spin-$1/2$ degrees of freedom, where dynamics under a classical Ising Hamiltonian is periodically alternated with a kick along the transverse direction.
The Floquet unitary describing a single evolution period of the kicked Ising model is given by 
\begin{align}\label{eq:U_Floq}
    U_F = e^{-i H_{\textrm{K}}} e^{-i H_\textrm{I}}\,,
\end{align}
generated by the Hamiltonians
\begin{align}
    H_{\textrm{I}} = J \sum_{j=1}^{L-1} \sigma^z_j \sigma^z_{j+1} +\sum_{j=1}^L h_j \sigma^z_j, \quad
    H_{\textrm{K}} =  b \sum_{j=1}^L \sigma_j^x.
\end{align}
Here $\sigma_j^{\alpha}$ with $\alpha \in \{x,y,z\}$ are the Pauli matrices, $L$ is the length of the chain, $J$ and $b$ are the Ising interaction strength and the transverse kick strength, respectively. $\{h_j\}$ describes a (possibly inhomogeneous) longitudinal field. We fix $J = b = \pi/4$, corresponding to dual-unitary dynamics~\cite{akila2016particle,bertini2018exact,bertini2019entanglement,gopalakrishnan2019unitary}. 

Despite its simplicity, the dynamics of Eq.~\eqref{eq:U_Floq} are generically ergodic and serve as a benchmark for chaotic quantum systems. Its space-time duality, identified in Ref.~\cite{akila2016particle}, was used to prove that its spectral form factor shows random matrix statistics, a fingerprint of quantum chaos~\cite{bertini2018exact}.
This model shows maximal entanglement growth~\cite{bertini2019entanglement,gopalakrishnan2019unitary,zhou_maximal_2022}, is scrambling~\cite{bertini_scrambling_2020,claeys_maximum_2020}, and its correlation functions decay to the ergodic values~\cite{claeys_maximum_2020,bertini2019exact}. 
Furthermore, this model was the first to provably exhibit `deep thermalization'~\cite{ho_exact_2022}, where Haar-random states locally emerge after projective measurements on subsystems. The proof interprets many-body dynamics as measurement-based quantum computation in the spatial direction~\cite{ho_exact_2022,stephen_universal_2024}, linking averaging over measurement outcomes to sampling from a universal gate set. We use this result to present a simple and exact derivation of the full delocalization dynamics in the Hilbert space, even at finite times, by relating sampling overlaps to sampling a Haar-random unitary matrix element in the spatial direction.

Dual-unitarity can be made explicit by introducing a diagrammatic notation
\begin{align}\label{eq:graphical_elements}
\vcenter{\hbox{\includegraphics[height=0.16\columnwidth]{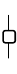}}} = 
\begin{pmatrix}
    1 & i \\
    i & 1
\end{pmatrix}, \qquad 
\vcenter{\hbox{\includegraphics[height=0.16\columnwidth]{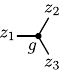}}} = \delta_{z_1 z_2 z_3}\, e^{ig(1-2z_1)},
\end{align}
where $z_j \in \{0,1\}$ label Fock states in the $\sigma^z$ basis and $g$ is an arbitrary phase. The Ising interactions and kicks can be graphically represented as~\footnote{Up to an irrelevant complex phase.}
\begin{align}
 e^{-i \frac{\pi}{4} \sigma^x} = \frac{1}{\sqrt{2}}\,\vcenter{\hbox{\includegraphics[height=0.16\columnwidth]{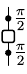}}}\,,\quad e^{-i \left(\frac{\pi}{4} \sigma_1^z \sigma_2^z + h_1 \sigma_1^z+h_2 \sigma_2^z\right)} = \vcenter{\hbox{\includegraphics[height=0.16\columnwidth]{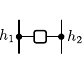}}}.
\end{align}
Dual-unitarity is made explicit by noticing that the phases from Ising interactions and kicks are interchangeable by swapping the roles of space (horizontal) and time (vertical)~\cite{claeys2022emergent,borsi2022construction,claeys_operator_2024,claeys2023dual,ho_exact_2022,stephen_universal_2024}. Note that this graphical language of delta tensors and Hadamard matrices closely relates to the ZX calculus~\cite{van_de_wetering_zx-calculus_2020}.

We consider the time evolution of an initial product state, $\ket{\psi(t=0)}=\ket{00\dots 0}$ as $\ket{\psi(t)} = U_F^t \ket{\psi(t=0)}$. The overlap of $\ket{\psi(t)}$ with a Fock state $\ket{z} = \ket{z_1 \dots z_L}$, with $z_i \in \{0,1\}$, can be graphically represented as
\begin{align}\label{eq:overlap_graphical}
\braket{z| \psi(t)} =  \frac{1}{2^{t L/ 2}}\,  \vcenter{\hbox{\includegraphics[width=0.50\linewidth]{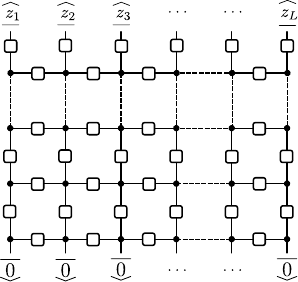}}},
\end{align}
where we made the phases implicit. The space-time duality is apparent: Exchanging space and time corresponds to rotating Eq.~\eqref{eq:overlap_graphical} by 90$^{\circ}$, leaving the bulk of the diagram invariant.

\emph{Delocalization from space-time duality.}--- To quantify Fock space delocalization, we define inverse participation ratios (IPRs)~\cite{wegner_IPR_1980,evers_IPR_2000} and corresponding participation entropies~\cite{baumgratz_quantifying_2014} as
\begin{align}
    I_q(t) = \sum_{z\in \mathcal{B}} |\langle z|\psi(t) \rangle|^{2q}, \quad  S_q(t) = \frac{1}{1-q} \ln[I_q(t)]\, ,
\end{align}
where $\ket{z} = \ket{z_1,z_2,\dots z_L}$ with $z_i \in \{0,1\}$ and $\mathcal{B}$ are bit-strings labelling these Fock states.
For a fully localized state, $I_q = 1$ and hence $S_q=0$. For a fully delocalized state that spans the entire space homogeneously, $|\langle z|\psi(t)\rangle| = 2^{-L/2}$ and $I_q = 2^{(1-q)L}$, resulting in a maximal value of $S_q = L \ln(2)$, which scales linearly with system size. 

Space-time duality allows us to rewrite the overlaps Eq.~\eqref{eq:overlap_graphical} as
\begin{align}\label{eq:factorization_overlap}
\braket{z_1 \dots z_L | \psi(t)} = \frac{1}{2^{L/2}} \bra{L} U(z_1) U(z_2) \dots U(z_L) \ket{R}\,,
\end{align}
where we have introduced unitary spatial operators acting on a $2^{\tau}$-dimensional Hilbert space with $\tau = t-1$, generating evolution in space as
\begin{align}
U(z) = \frac{1}{2^{\tau/2}}\,\, \vcenter{\hbox{{\includegraphics[height=0.22\columnwidth]{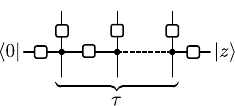}}}}\,, 
\end{align}
here rotated by 90$^\circ$ for convenience, and two boundary vectors 
\begin{align}
\bra{L} = \frac{1}{2^{\tau}}\,\,\vcenter{\hbox{\includegraphics[height=0.2\columnwidth]{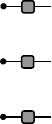}}}\,\,, \qquad
\ket{R} = \,\,\vcenter{\hbox{\includegraphics[height=0.2\columnwidth]{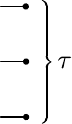}}}\,\,.
\end{align}
The gray square represents the complex conjugate of Eq.~\eqref{eq:graphical_elements}, $\vcenter{\hbox{\includegraphics[height=.5\baselineskip]{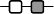}}} = 2 \,\,\vcenter{\hbox{\includegraphics[height=.5\baselineskip]{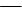}}}$.
Eq.~\eqref{eq:overlap_graphical} has a straightforward interpretation. Rather than considering the dynamics in time (along the vertical direction) of a state in a $2^L$-dimensional Hilbert space, we can consider the dynamics in space (horizontal direction) of a state in a $2^\tau$-dimensional Hilbert space. The overlap $\braket{z_1 z_2 \dots z_L|\psi(t)}$ follows as the appropriate matrix element of the product of unitary matrices $U(z_1) U(z_2) \dots U(z_L)$.

Our proof follows similar arguments in Ref.~\cite{ho_exact_2022} on deep thermalization: The two unitary matrices \( U(0) \) and \( U(1) \) constitute a universal gate set, such that in the thermodynamic limit \( L \to \infty \), the products \( U(z_1) U(z_2) \dots \) uniformly sample the space of unitaries~\footnote{In order to constitute a universal gate set, the phases need to be chosen such that these gates are tuned away from Clifford points, which here corresponds to choosing $g \neq \mathbbm{Z} \pi/8$ (see Ref.~\cite{ho_exact_2022}).}. As a result, in the large \( L \) limit, the summation over Fock states \(\ket{z}\) and the corresponding unitaries can be replaced by the average over Haar-random unitary matrices~\cite{ho_exact_2022}, giving
\begin{align}\label{eq:IPR_to_ULR}
     I_q = \sum_{z \in \mathcal{B}} |\langle z|\psi(t) \rangle|^{2q} = 2^{L(1-q)} \times \mathbbm{E}_{\textrm{Haar}}\left[|U_{LR}|^{2q}\right],
\end{align}
where the expectation value refers to averaging over Haar-random matrices $U$ with dimension $d=2^{\tau}$ and $U_{LR} = \langle L |U|R\rangle$. $\mathbbm{E}_{\textrm{Haar}}\left[|U_{LR}|^{2q}\right]$ can be computed using the Haar-random unitary toolbox~\cite{weingarten_asymptotic_1978}\footnote{See Supplemental material for more details.\label{SM}}, resulting in
\begin{align}\label{eq:IPR_DU}
    I_q(\tau+1) =2^{L(1-q)}\frac{q! \, 2^{q\tau}}{2^\tau (2^\tau+1)\dots (2^\tau+q-1)}\,,
\end{align}
with corresponding participation entropies
\begin{align}\label{eq:S_q_DU}
    &S_q(\tau+1) = L \ln(2) \nonumber\\
    &\qquad \qquad+ \frac{1}{1-q} \ln\left[\frac{q!\, 2^{q \tau}}{2^\tau (2^\tau+1)\dots (2^\tau+q-1)}\right]\,.
\end{align}
Two immediate limits are clear: At $t=1$, the state is maximally delocalized and $S_q(t=1) = L \ln(2)$, since $\ket{\psi(t=1)} = \otimes \frac{\ket{0} + i \ket{1}}{\sqrt{2}}$ such that \(|\langle z|\psi(t=1)\rangle|^2 = 2^{-L}\). 
As $t \to \infty$, we recover $S_q(t \to \infty) = L \ln(2) + \ln(q!)/(1-q)$, corresponding to the IPRs for a Haar-random state.
Eqs.~\eqref{eq:IPR_DU},~\eqref{eq:S_q_DU} fully characterize the dynamics of the IPRs and participation entropies and compose the first result of this work. Interestingly, the convergence to the Haar-random value happens on a time scale $t \propto \ln(q!)$, independent of system size. 

\begin{figure}[t]
  \includegraphics[width=\columnwidth]{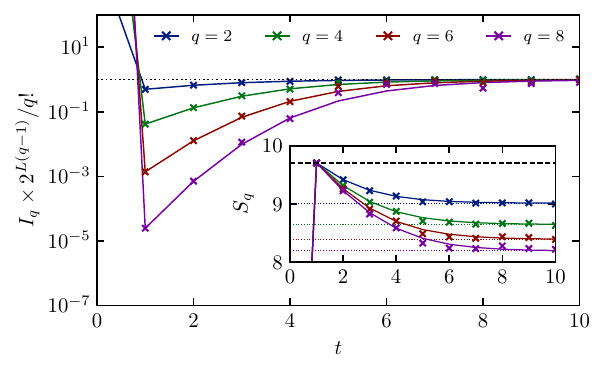}
  \caption{Inverse participation ratios $I_q$ and the corresponding participation entropies $S_q$ as a function of time for $q=2, 4, 6, 8$, with $g=\pi/3$ and system size $L=14$. The IPRs ($I_q$) are normalized by the Haar random value $q!/2^{L(q-1)}$. Both rapidly converge to their steady-state value, indicated by a dashed line, where crosses indicate numerics and full lines analytical results. 
  \label{fig:IPR}}
\end{figure}

Even though our results are exact only for $L \rightarrow \infty$, we obtain a good agreement with finite size numerics already for $L=14$, as shown in Fig.~\ref{fig:IPR}, which displays $I_q$ and $S_q$ for several values of $q$. All phases in Eq.~\eqref{eq:overlap_graphical} are chosen as $g = \pi/3$ for concreteness. 
We emphasize that these results hold without averaging and for homogeneous models in both space and time, although they also apply when all phases are chosen differently.

The probability distribution of Fock-state probabilities, i.e. $\mathcal{P}(p = |\langle z|\psi(t) \rangle|^2)$, is known as the distribution of output bit-string probabilities. For a Haar random state this distribution is the Porter-Thomas (exponential) distribution, $\mathcal{P}(p) \propto e^{-Np}$, where $N$ is the Hilbert space dimension. From $I_q$ in Eq.~\eqref{eq:IPR_DU} it is possible, in principle, to reconstruct the full distribution from its moments, since 
\begin{align}
I_q = \sum_{z \in \mathcal{B}} |\langle z |\psi \rangle|^{2q} = 2^L \int_{0}^1 dp \, \mathcal{P}(p)\, p^q\,.
\end{align}
However, it is more direct to observe that the IPRs are consistent with the moments of
\begin{align}\label{eq:P_p}
    \mathcal{P}_{\textrm{DU}}(p;t) = N (1-2^{-\tau}) \theta(2^\tau-Np)\left(1-\frac{N p}{2^\tau}\right)^{2^\tau-2},
\end{align}
where $N=2^L$ and $\theta$ is the Heaviside function. We can interpret this distribution as an appropriately rescaled Porter-Thomas distribution for a smaller Hilbert space set by the number of time steps, writing $\tilde{p} = Np$,
\begin{align}
    \mathcal{P} \propto \theta(2^\tau-\tilde{p}) \left(1-\frac{\tilde{p}}{2^\tau}\right)^{2^\tau-2},
\end{align}
consistent with Eq.~\eqref{eq:IPR_to_ULR}.

\begin{figure}[t]
  \includegraphics[width=\columnwidth]{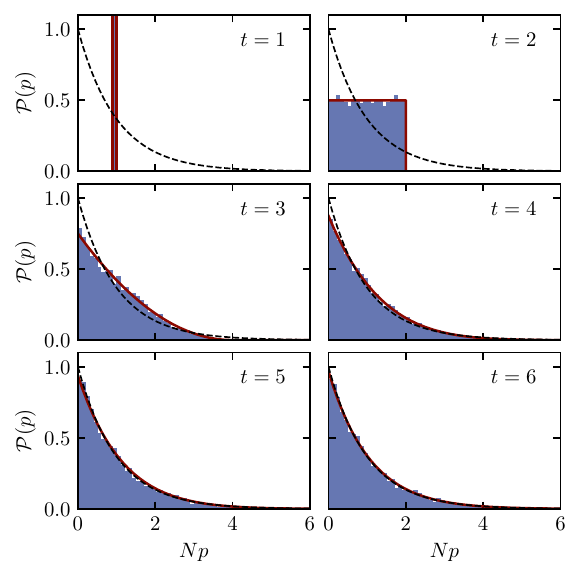}
  \caption{Distribution of bit-string probabilities/overlaps $p = |\langle z|\psi(t) \rangle|^2$ at different times. Numerical results (histograms) are compared with analytical predictions (full red lines), showing excellent agreement. After $t=6$ discrete time steps, the distribution is visually indistinguishable from the Porter-Thomas distribution (dashed line). Numerics performed for the same model as in Fig.~\ref{fig:IPR}.
  \label{fig:distribution}}
  \end{figure}

The full dynamics of this distribution presents the second main result of this work. This distribution also exhibits interesting transient dynamics. 
At $t=1$, $\mathcal{P}(p)$ is delta-distributed since $|\langle  z|\psi(t=1) \rangle|^2=1/2^L$, at $t=2$ we get a uniform distribution in the interval $[0,1/2^{L-1}]$. Afterward, $\mathcal{P}(p)$ approaches the Porter-Thomas distribution exponentially fast as time increases. Fig.~\ref{fig:distribution} shows the comparison between the analytical prediction and finite-size numerics, observing excellent agreement and the rapid emergence of the Porter-Thomas distribution.

\emph{Stability to local perturbations.}---We now consider the stability of our results away from dual-unitarity. We focus on a simple class of perturbations, locally breaking dual-unitarity at the boundary, for which calculations remain analytically tractable. The resulting dynamics are expected to be representative for generic local perturbations to dual-unitarity.
In the unitary dynamics of Eq.~\eqref{eq:overlap_graphical}, the single-site unitary acting on the last site is replaced by a generic single-site unitary $u \in U(2)$, such as a `detuned' kick $u = e^{-i \theta \sigma^x}$ with $|\theta| \neq \pi/4$. A similar setup was considered in Ref.~\cite{kos_thermalization_2021} to study thermalization dynamics and spectral statistics, establishing quantum chaotic dynamics in a large class of models. 

The evolved state can be diagramatically represented as in Eq.~\eqref{eq:overlap_graphical} where we replace the rightmost gate by a generic unitary matrix $u$, represented (with its complex conjugate) as
\begin{align}
u = \frac{1}{\sqrt{2}}\,\,\vcenter{\hbox{\includegraphics[height=0.1\columnwidth]{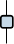}}}\,\,,\qquad
u^* = \frac{1}{\sqrt{2}}\,\,\vcenter{\hbox{\includegraphics[height=0.1\columnwidth]{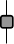}}}\,\,.
\end{align}
This local perturbation can be absorbed in our derivation by modifying the right boundary vector to
\begin{align}
  \ket{R(z)}= \frac{1}{2^{\tau/2}}\,\vcenter{\hbox{\includegraphics[width=0.45\columnwidth]{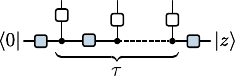}}}\,, 
\end{align}
again rotated by 90$^\circ$ for convenience. The above derivation can be directly repeated, where we only need to separate the summation over $z_L$ to return
\begin{align}
    I_q(\tau+1) = \, &\frac{q!\, 2^{L(1-q)}}{2^\tau (2^\tau+1)\dots (2^\tau+q-1)} \nonumber\\
    &\quad \times \frac{1}{2}\sum_{z \in \{0,1\}} |\langle R^*(z)|R(z) \rangle|^q.
\end{align}
That this expression only depends on the boundary through its norm can be understood since the Haar measure is invariant under unitary transformations, such that when evaluating the matrix elements w.r.t. a state only the norm of this state will matter. The norm of $\ket{R(z)}$ can be expressed as
\begin{align}
   \ \vcenter{\hbox{\includegraphics[width=0.7\columnwidth]{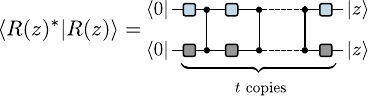}}}\,\,,
\end{align}
where the phases in the delta tensors cancel. We write $\braket{R(z)^*|R(z)}= 2^t \bra{0}\mathcal{M}^t \ket{z}$, where $\mathcal{M}$ is a unistochastic matrix defined as $\mathcal{M}_{ij} = |u_{ij}|^2$. We can rewrite the expression for the IPR as
\begin{align}
    I_q(t) = I_q^{\textrm{DU}} (t) \times 2^{q-1} \sum_{z \in \{0,1\}} \langle 0|\mathcal{M}^t|z\rangle^q\,.
\end{align}
where we have factored out the dual-unitary result [Eq.~\eqref{eq:IPR_DU}].
\begin{figure}[t]
  \includegraphics[width=\columnwidth]{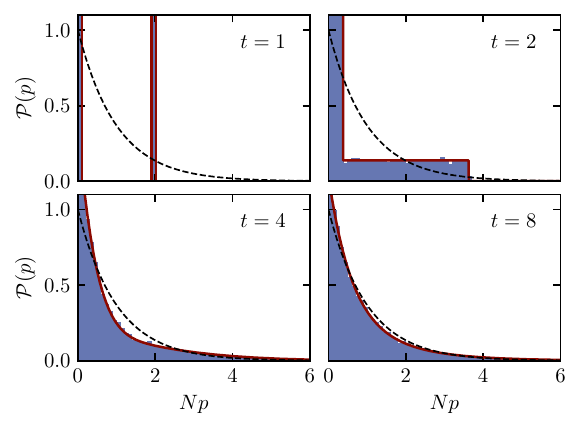}
  \caption{Dynamics of the distribution of bit-string probabilities/overlaps $p = |\langle z|\psi(t) \rangle|^2$ for dual-unitary dynamics with a local perturbation. Numerical results (histograms) are compared with analytical predictions [Eq.~\eqref{eq:dist_PT}] (full red lines), showing excellent agreement and a slower convergence to the Porter-Thomas distribution (dashed lines). Numerics performed for the same model as in Fig.~\ref{fig:IPR} and $\theta = \pi/14$.
  \label{fig:distribution_PT}}
  \end{figure}

For a `generic' $u$, i.e., all matrix elements are nonzero, the Perron–Frobenius theorem guarantees that the matrix $\mathcal{M}$ has a unique leading eigenvalue $1$ with left and right eigenstates given by $\ket{+} = (\ket{0}+\ket{1})/\sqrt{2}$. We hence find that
\begin{align}
    \lim_{t \to \infty} \bra{0}\mathcal{M}^t \ket{z} = \langle 0|+\rangle \langle +|z\rangle =\frac{1}{2},
\end{align}
and we recover the Haar random result
\begin{align}
    \lim_{t \to \infty} I_q(t) = q!\,2^{L(1-q)}\,.
\end{align}
The corresponding overlap distribution can be obtained from the IPRs, and its long-time limit will converge to the Porter-Thomas distribution. For concreteness, we consider a representative gate $u=e^{-i \theta \sigma^x}$, for which $\mathcal{M}$ can be directly diagonalized to return
\begin{align}\label{eq:IPR_PT}
    I_q(t) = I_{q}^{\textrm{DU}}(t)\times \frac{[1+\cos^t(2\theta)]^q+[1-\cos^t(2\theta)]^q}{2}\,.
\end{align}
Defining $c_{\pm}(t) = 1 \pm \cos^t(2\theta)$, these IPRs correspond to the moments of the distribution
\begin{align}\label{eq:dist_PT}
    \mathcal{P}(p;t) = \frac{1}{2}\Bigg[ \frac{\, \mathcal{P}_{\textrm{DU}}\left({p}/{c_+(t)};t\right)}{c_+(t)}+\frac{\, \mathcal{P}_{\textrm{DU}}\left({p}/{c_-(t)};t\right)}{c_-(t)}\Bigg]\,.
\end{align}
In the limit $\theta = \pi/4$ we recover the expected dual-unitary result since $c_{\pm}(t) = 1$. Otherwise, the time scale to reach the dual-unitary result follows as $t \propto 1/\ln|\cos(2\theta)|$. The local perturbation effectively acts as a bottleneck in the approach to the Porter-Thomas limit. For $\theta=0$ we never approach this limit, since the local perturbation reduces to the identity and $z_L = 0$ at all times, resulting in nonergodic dynamics. We compare the expression for the distribution with numerics in Fig.~\ref{fig:distribution_PT}, again observing excellent agreement. The local perturbation results in a slower approach to the Porter-Thomas distribution and $\mathcal{P}(p;t)$ is generally skewed towards smaller probabilities.

These constitute the third main result of this work: We exactly characterize Fock-space delocalization in the presence of a generic perturbation that locally breaks dual-unitarity and establishes the appearance of Haar-random states, albeit at a time scale that is parametrically longer than in the dual-unitary case. Even though the effect of the perturbation on the dynamics is highly nontrivial, the resulting probability distribution can still be straightforwardly obtained. 

\emph{Conclusions $\&$ outlook}.--- In this work, we examine the spreading of an initial product state in the Fock space for dual-unitary dynamics, focusing on the self-dual kicked Ising model, a paradigmatic example of dual-unitarity and chaotic dynamics. We analytically characterize the spreading by computing the inverse participation ratios and the full overlap probability distribution over time. The initial state delocalizes exponentially fast in the Fock space, rapidly reaching a steady state described by Haar random states with the overlap probability distribution given by the Porter-Thomas distribution. This process occurs on a time scale independent of system size, unlike in random quantum circuits where it scales logarithmically with system size. We also investigate the stability of our results under local perturbations that break dual-unitarity, finding that these slow down the spreading in the Fock space. 

These results further cement the self-dual kicked Ising model as a `maximally chaotic' model in which many-body quantum chaos can be analytically characterized -- consistent with previous observations on its spectral properties~\cite{bertini2018exact}, entanglement spreading~\cite{bertini2019entanglement,gopalakrishnan2019unitary}, and deep thermalization~\cite{ho_exact_2022}. Furthermore, this motivates the use of dual-unitary circuits for the preparation of random states in e.g. quantum simulation and tomography~\cite{huang_predicting_2020,richter_simulating_2021,mcginley_shadow_2023,mark_benchmarking_2023}.
There are various natural extensions of our work. The kicked Ising model can be extended either to higher dimensions or to different lattice structures, which also allow for the introduction of kinetic constraints~\cite{mestyan_multi-directional_2022,liu2023solvable,osipov_correlations_2023,rampp_entanglement_2023,sommers_zero-temperature_2024}. Relating the Fock-space delocalization to entanglement growth and operator spreading, and investigating local measurements to allow entanglement transitions, remain objectives for future study.

\textit{Acknowledgments.}---We thank Jiangtian Yao for illuminating discussions. G.~D.~T. acknowledges the support from the EPiQS Program of the Gordon and Betty Moore Foundation.

\bibliography{Library}

\begin{widetext}
\newpage
\appendix

\section{Supplemental material of ``Fock-space delocalization and the emergence of the Porter-Thomas distribution \\
from dual-unitary dynamics''}

\subsection{Averaging over Haar-random unitary matrices}
In this we give more detail on the derivation of $I_q$ and in particular the expectation value of Eq.~\eqref{eq:IPR_to_ULR}, which involves the computation of the average $U_{LR}$ over Haar-random unitaries.  We write down the expectation value of 
$\mathbbm{E}_{\textrm{Haar}}\left[|U_{LR}|^{2q}\right] = 
\mathbbm{E}_{\textrm{Haar}}\left[|\langle L|U|R \rangle|^{2q}\right]$
where $U$ are Haar-random matrices with dimension $d=2^{\tau}$ and boundary vectors are defined in the main text,
as
\begin{align}
\mathbbm{E}_{\textrm{Haar}}\left[|U_{LR}|^{2q}\right] = \bra{\tilde{L}_q} T_q \ket{\tilde{R}_q},
\end{align}
where we consider $q$ replicas of $U$ and $U^*$ as
\begin{align}\label{eq:transfermat}
   T_q =  \mathbbm{E}_{\textrm{Haar}}[\underbrace{U \otimes U^* \otimes \dots  U \otimes U^*}_{q \, \textrm{replicas}}] \,,
\end{align}
and corresponding replicated boundary vectors
\begin{align}
\bra{\tilde{L}_q} = \underbrace{\bra{L} \otimes \bra{L^*} \otimes \dots  \bra{L} \otimes \bra{L^*}}_{q \, \textrm{replicas}}\,, \qquad
\ket{\tilde{R}_q} = \underbrace{\ket{R} \otimes \ket{R^*} \otimes \dots  \ket{R} \otimes \ket{R^*}}_{q \, \textrm{replicas}}\,.
\end{align}
The expectation value Eq.~\eqref{eq:transfermat} is a standard workhorse of random matrix theory and can be expressed in terms of the Weingarten functions~\cite{weingarten_asymptotic_1978}
\begin{align}
T_q  = \sum_{\sigma,\tau \in \Sigma_q} \textrm{Wg}(\sigma \tau^{-1},d=2^{\tau}) \ket{\tau_q} \bra{\sigma_q}\,,
\end{align}
where the sum runs over the permutations group $\Sigma_q$ of the $q$ replicas. Each permutation $\sigma\in \Sigma_q$ defines a state in the space of replicas as
\begin{align}
    \braket{i_1, i_1', \dots i_{q}, i_{q'}|\sigma} = \delta_{i_1, i'_{\sigma(1)}} \delta_{i_2, i'_{\sigma(2)}} \dots  \delta_{i_q, i'_{\sigma(q)}}\,,
\end{align}
with each index $i$ labelling a state in the $d=2^\tau$-dimensional Hilbert space. The Weingarten functions additionally satisfy
\begin{align}
    \sum_{\sigma \in \Sigma_q} \textrm{Wg}(\sigma,d=2^{\tau}) = \frac{1}{2^{\tau}(2^{\tau}+1)\dots (2^{\tau}+q-1)}\,.
\end{align}
It is straightforward to check that $\braket{\tilde{L}_q|\sigma} =1$ and $\braket{\sigma|\tilde{R}_q} = 2^{q \tau}$, $\forall \sigma \in \Sigma_q$, such that combining the above expressions directly returns the result from the main text,
\begin{equation}
\mathbbm{E}_{\textrm{Haar}}\left[|U_{LR}|^{2q}\right]= \sum_{\sigma,\tau \in \Sigma_q} \textrm{Wg}(\sigma \tau^{-1},d=2^{\tau}) \braket{\tilde{L}_q|\tau_q} \braket{\sigma_q|\tilde{R}_q} =\frac{q! \, 2^{q\tau}}{2^\tau (2^\tau+1)\dots (2^\tau+q-1)}.
\end{equation}
\newpage

\subsection{Comparison between Dual-Unitary and Random Circuits}
For completeness, in this section, we present an additional comparison between the self-dual Ising model, random unitary circuits, and perturbed dual-unitary models.

\begin{figure}[t]
  \includegraphics[width=\textwidth]{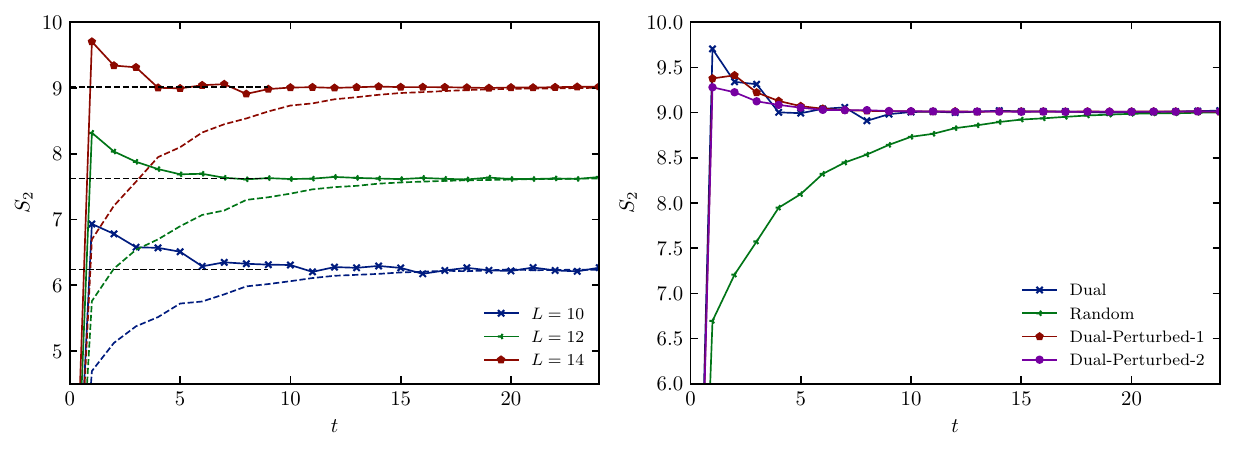}
  \caption{The left panel shows the participation entropy $S_2(t)$ as a function of time for several system sizes $L$. The solid lines represent the self-dual kicked Ising case, while the dashed lines represent the random unitary case.  The right panel displays $S_2(t)$ for a fixed $L=14$ and several models: the self-kicked Ising (Dual), random unitary circuits (Random), and two self-dual kicked Ising models perturbed locally (Dual-Perturbed-1,2). The black dashed lines are a guide for the eye and indicate the ergodic value $S_2(t\rightarrow \infty) = (L-1)\log{2}$. For the cases involving randomness, i.e., random gates, we average over 50 random instances.
  \label{fig:Comparison}}
\end{figure}

The left panel of Fig.~\ref{fig:Comparison} shows the participation entropy $S_2$ for the self-dual kicked Ising model (solid lines) and random unitary circuits (dashed lines) for several system sizes. Both models reach their ergodic value $S_q(t\rightarrow \infty) = L\log{2} +\log{q!}/(1-q)$ at late times; however, as described in the main text, the convergence to the equilibrium value is faster in the dual-unitary cases. Furthermore, as one can notice, for the random unitary case the time of this approach shifts with increasing system size, as predicted in Ref.~\cite{turkeshi_hilbert_2024}.

To further support our results, we investigate the effect of local perturbations that break dual-unitarity away from the boundary. The right panel of Fig.~\ref{fig:Comparison} shows the dynamics of $S_2(t)$ for a fixed system size $L=14$ across several models: the self-kicked Ising model (Dual), the random unitary circuit (Random), and two self-dual kicked Ising models perturbed locally (Dual-Perturbed-1,2). In the first perturbed model, we insert a random unitary gate at the central site, while in the second, we apply two-site random gates in the middle, changing them over time. The self-kicked model and the perturbed ones are the first to reach their ergodic values, followed later by the random unitary case. These numerical results provide further evidence for the stability of our results against local perturbations.

\end{widetext}

\end{document}